# Consecutive Insulator-Metal-Insulator Phase Transitions of Vanadium Dioxide by Hydrogen Doping

Shi Chen[1+], Zhaowu Wang[2, 3+], Lele Fan[1], Yuliang Chen[1], Hui Ren[1], Heng Ji[4], Douglas Natelson[4], Yingying Huang[5], Jun Jiang[2]*, ChongwenZou[1]*

[1]National Synchrotron Radiation Laboratory, University of Science and Technology of China, Hefei, 230026, China.

[2] Hefei National Laboratory for Physical Sciences at the Microscale, iChEM (Collaborative Innovation Center of Chemistry for Energy Materials), Hefei Science Center (CAS), and School of Chemistry and Materials Science, University of Science and Technology of China, Hefei, 230026, China.

[3] School of Physics and Engineering，Henan University of Science and Technology，Luoyang, Henan 471023, P. R. China

[4]Department of Physics and Astronomy, MS 61, Rice University, 6100 Main Street, Houston, Texas 77005, United States

[5]School of Physics and Optoelectronic Technology and College of Advanced Science and Technology, Dalian University of Technology, Dalian 116024, China

Corresponding Authors: jiangj1@ustc.edu.cn ; czou@ustc.edu.cn

+These two authors contributed equally to this paper.

# Abstract

We report modulation of a reversible phase transition in $VO_2$ films by hydrogen doping. Metallic and insulating phases are successively observed at room temperature as the doping concentration increases. The polarized charges induced by doped hydrogens gradually occupy V*3d*-O*2p* hybridized orbitals and consequently modulate the filling of the $VO_2$ crystal conduction band-edge states, which eventually become saturated and therefore evolve into new valence band-edge states. A clear linear relationship between H-concentration and V3d orbital occupancy were found. These results suggest the exceptional sensitivity of $VO_2$ electronic properties to electron concentration and orbital occupancy, providing key information for the phase transition mechanism.

Since Mott predicted the first-order metal-insulator transition (MIT) behavior several decades ago, the underlying physics of how electron-electron interactions affect electronic structures of strongly correlated systems has become a hot field of condensed matter physics. As a typical correlated oxide, monoclinic vanadium dioxide (M-$VO_2$) undergoes a MIT process from metal to insulating state at the critical temperature near 68$^oC$ [1,2]. This transition is believed to be driven by the Mott transition associated with electron-electron correlations [3–5] or the Peierls transition involving electron-phonon interactions [6,7], although the actual MIT mechanism is still under debate.

Regulating electronic density through atomic doping is an effective way to modulate the balance between competing phases in strongly correlated materials [8,9]. It is reported that hydrogen incorporated into M-$VO_2$ crystals can stabilize the metallic $VO_2$ phase at room temperature [10-14] and further lead to an insulating phase in the high doping regime [15]. Fig. 1(a) shows both the normal temperature-triggered MIT (from insulating M-$VO_2$ to metallic rutile $VO_2$ (R-$VO_2$)) as well as the phase transitions driven by reversible hydrogen doping. Different from the temperature induced MIT, the metallic H-$VO_2$ (H-$VO_2$(M)) is produced by light doping concentration in $H_xVO_2$ ($0< x < 1$), while further hydrogenation treatment up to the saturation point of $HVO_2$ leads to a new insulating H-$VO_2$ (H-$VO_2$(I)) state.

Explained in the context of crystal-field theory [16,17], the H-doping induced band structure evolutions is represented in Fig. 1(b). For the pure M-$VO_2$ crystal, the Fermi level is located in the ~0.6eV insulating gap between the edge states of $d_{//}$ and $d_{//}^*/\pi^*$ orbitals. Under low H-doping treatment, the conduction band-edge states of $d_{//}^*/\pi^*$ are partially filled, resulting in a metallic state. Upon further hydrogenation, the $d_{//}^*$ band become fully occupied, producing new valence band-edge states. This leads to a new insulating gap and an insulating H-doped $VO_2$ crystal. As electron-electron correlations and electron-lattice interactions are always relevant, the variation of electronic structure also influences the crystal lattice structure, such as the V-V bonds. However, largely due to the difficulties of characterizing the dynamic doping process from incompletely hydrogenated states to the maximally H-doped state, previous work focused on the phenomena of doping-induced property variations, and thereby did not arrive at a clear mechanism for the doping-based transitions.

In this letter, we report in-situ synchrotron based characterization to directly monitor the dynamics of H atom doping in $VO_2$ crystals. This characterization captures the lattice/orbital variations as well as the band structure evolution during the reversible phase transitions. Based on the experimentally determined geometric details, first-principle theoretical studies show the gradual filling of conduction band-edge states with nearly linear dependence on H-doping level, which was unattainable in previous works. The underlying mechanism for the doping-driven insulator-metal-insulator phase transitions is thus revealed, indicating the filling-based evolution of the semiconductor conduction band-edge states to valence band-edge states.

The 50 nm M-$VO_2$/$Al_2O_3$ (0001) epitaxial films were deposited by an rf-plasma assisted oxide-MBE method [18]. Metallic H-$VO_2$ (M) and insulated H-$VO_2$ (I) films were obtained by hydrogenation for different time in Ar/15% $H_2$ gas mixture with Au particles as the catalyst [12, 15]. From the in-situ R-T measurement in vacuum conditions, the insulating H-$VO_2$ (I) gradually became a metallic sample as shown in Fig.1 (c). This metallic state was quite stable in vacuum. Upon further annealing of this metallic H-$VO_2$ (M) sample in air, the material will gradually revert to its initial monoclinic structure as shown in Fig.1 (d). In fact, if conducting the R-T tests in air, the phase transformation of the hydrogenated $VO_2$ film from the insulating state to metallic state, and then back to the initial (monoclinic) insulator state could be observed (Fig.S1 in Ref. [19]). This suggests that the hydrogenated $VO_2$ film is more stable in vacuum than in the air condition.

The chemical states of the hydrogenated $VO_2$ were examined by X-ray photoelectron spectroscopy (XPS) in Fig.2 (a). There exist double peaks for V2p3/2 at 516.0 eV and 517.1 eV, which are assigned to $V^{4+}$ and $V^{5+}$ states, respectively [20,21]. The O*1s* peak at 530.1 eV is from vanadium oxide and the peak at 531.9 eV originates from the –OH species [12,22,23]. For hydrogenated $VO_2$, the -OH peak is much stronger than V-O peak, indicating the formation of -OH bonds [22,23]. After annealing the H-$VO_2$ samples in air for 30 min, the –OH peak decreases substantially and the sample almost reverts to the undoped M-phase. We can then estimate the hydrogen concentration by XPS peak fitting (Fig.S2 in Ref. [19]) and obtain the stoichiometric values of $H_{0.64}VO_2$ for the metallic H-$VO_2$ (M) and $H_{0.89}VO_2$ for the insulating H-$VO_2$ (I). Nevertheless, it is difficult to determine the exact H-atom concentration because of possible intermediate phases during experimental measurements.

The evolution of electronic structures can be inferred from measuring optical band gap. In-situ UV-Vis spectra in Fig.2 (b) indicated the enhancement of visible transparency along with increasing H-concentration in $VO_2$, suggesting the gradual decreasing/filling of empty $d_{//}$*sates at conduction band edge [24, 25]. Meanwhile, synchrotron-based X-ray diffraction (XRD) spectra in Fig. 2 (c) showed that diffraction peaks of hydrogenated $VO_2$ shifted towards low angles, indicating the presence of slightly enlarged lattice parameters and expanded cell volumes.

Synchrotron based XANES spectra suggested that H atoms intercalation actually caused electron doping effect. The V L-edge curves continuously shifted to lower energies as the H concentration increases (Fig.S3 in Ref. [19]), implying charge transfer to V atoms that caused valence state reduction from $V^{+4}$ to $V^{+(4-\delta)}$, or even to the $V^{+3}$ state [12]. The O K-edge mapping in Fig.2 (d) presented the transitions from O1s to O2p states, in which two distinct features of the $t_{2g}$ and $e_g$ peaks reflected the unoccupied states [26, 27]. The intensity ratio of $t_{2g}/e_g$ decreased substantially with increasing H concentration, meaning that $t_{2g}$ levels including the $d_{//}$* and $\pi$* orbitals were gradually filled by electron doping. Considering the valence band structures shown in Fig.1 (b), the doped electrons will occupy the lower $t_{2g}$ levels, resulting in an up-shift of the Fermi level. As the $t_{2g}$ levels were gradually filled by electron doping, the intensity of $t_{2g}$ peak decreased gradually, while the higher-energy $e_g$ peak increased simultaneously.

The above characterizations thus provide structural details for a thorough theoretical investigation on the microscopic kinetics of phase transitions. First-principles simulations at the density function theory (DFT) level were applied to examine the H-doping induced geometric and electronic structure variations [19]. Fig.3 shows $VO_2$ models with corresponding calculation results. Compared to the pure monoclinic $VO_2$ (1×1×1) unit cell ($V_4O_8$), the optimized structures of both lightly H-doped cells ($HV_4O_8$ or $H_2V_4O_8$) and saturated doping cells ($H_4V_4O_8$) are quite stable with negative formation energies and H-O bonds of ~1.00 Å [19]. H-doping induces only subtle geometric variations including slightly enhanced lattice parameters and expanded cell volumes, consistent with above characterizations and previous reports [10]. In contrast, the electronic structures were heavily affected by H-doping. All hydrogen atoms hold strong positive charges (> 0.64 $e^+$), inducing polarized electrons in V-O bonds. The energy distribution of polarized electrons can be inferred from the simulated density of states (DOS). For the pure $V_4O_8$ cell, valence band edge states are well separated from the conduction band in Fig.3 (d), accounting for the insulating properties of a typical semi-conductor band structure (Fig. 3(g)). With one or two H introduced into $V_4O_8$, the $HV_4O_8$ (Fig.S4 in Ref [19]) or $H_2V_4O_8$ cells exhibit partially occupied conduction band-edge states. The newly occupied states are mostly $d_{//}$* orbitals of V atoms, suggesting that H-doping induce more d-electrons. The calculated charge densities clearly show the electron distributions and the charge transfer for the H-doped system (Fig.S5 in Ref. [19]). The polarized d-electrons then shift up the Fermi level, and consequently cause the conduction band edge in $V_4O_8$ (Fig.3 (g)) to become partially occupied (Fig.3h) for $H_2V_4O_8$, naturally leading to an insulator-to-metal transition. The evolution of electronic structures becomes more interesting when we add 4 H atoms into a $V_4O_8$ cell. The computed DOS in Fig.3 (f) shows that the polarized electrons fully occupy the original conduction band edge states in $V_4O_8$. These states with lowered energies become the new valence band maximum (Fig. 3i), leading to an insulating ground state. The effect of varying different interstitial H sites was tested, giving similar electronic structures (Fig.S6 in Ref. [19]).

In order to unravel the actual microscopic mechanism, the gradual filling of V-3d orbitals with increasing H-concentration were simulated. A larger (1×2×1) cell of $V_8O_{16}$ was constructed. These together with the above H-doped $V_4O_8$ models enabled us to examine the effect of varying doping concentration ($H_xV_8O_{16}$ with x= 0~8). The DOS features in Fig. 4(a) and Fig. S7 show that increasing H-doping gradually shifted up the Fermi level to exceeding the conduction band edge state of the V-3d orbitals [19]. The V-3d occupancy was computed by integrating the corresponding DOS below the Fermi level (Fig. 4(a)). Surprisingly, the V-3d occupancy increases linearly with H-doping concentration (Fig. 4(b)). This clear relationship thus explains the phase transitions, from the insulating state ($VO_2$) with empty V-3$d_{//}$* orbitals, to the metallic states ($H_xV_8O_{16}$, x=1~7) with un-saturated H-doping and partially-occupied conduction band edge states, and eventually to the new insulating state ($H_8V_8O_{16}$) with saturated H-doping and fully-occupied V-3$d_{//}$* orbitals which already became the new valence band edge. It should be noticed that the electron correlations in $VO_2$ system need to be accounted by many-body techniques. To test the consistency, both PBE+U and advanced HSE06 computations were conducted and exhibited nearly the same linear filling of V-3$d_{//}$* orbitals by H-doping (Fig.S8-9 in Ref. [19]), providing key insights for mechanism investigation.

In conclusion, the present experimental and theoretical work provides a comprehensive description of the whole dynamic process of hydrogen-induced consecutive phase transitions in $VO_2$ crystal. We accomplished direct detection of gradual electron occupation of the $e_g/t_{2g}$ orbitals along with increasing H-concentration, when the formation of O–H bonds is accompanied by electrons donating from H to V and O atoms. These polarized electrons cause the partial filling of the conduction band edge, leading to metallic behavior. With saturated O–H bonds, the additional polarized electrons eventually occupy all $d_{//}$ orbitals at the previous conduction band edge, creating new valence band edge states and producing a new insulating gap. A linear relationship describing the dependence of $d_{//}$ orbital occupancy on H-doping concentration was revealed. These data and analysis provide fundamental physical insights into the H-doping-induced electronic band/orbital occupancy variations and reveal the underlying mechanism of the reversible doping controlled insulator-metal-insulator phase transition. This approach should lead to the accurate control of electronic properties of this and other oxide materials in a continuously adjustable manner.

This work was partially supported by the National Basic Research Program of China (2014CB848900), the National Natural Science Foundation of China (U1432249, 11574279, 11404095), the Fundamental Research Funds for the Central Universities and the research foundation of Key Laboratory of Neutron Physics, China Academy of Engineering Physics (Grant No. 2013BB04). The authors also acknowledge support from the Beijing Synchrotron Radiation Facility, Shanghai Synchrotron Radiation Facility and the National Synchrotron Radiation Laboratory (NSRL) of Hefei. DN and HJ acknowledge support from U.S. DOE Office of Science/Basic Energy Sciences award DE-FG02-06ER46337.

## Figure Captions

FIG. 1 (a) Scheme for the phase transitions induced by temperature and electron doping. For the hydrogenation effect, two hydrogen-doped $VO_2$ phases exist, which show insulating (H-$VO_2$ (I)) and metallic (H-$VO_2$ (M)) states, respectively; (b) Molecular orbitals and band diagrams for monoclinic $VO_2$ insulator and H-doped $VO_2$ in its metallic/insulating phase; (c) R-T curves for metallic H-$VO_2$ (M) and insulating H-$VO_2$ (I) film samples. During a series of 20 heating/cooling cycles in vacuum over the range of 35 $^{o}$C to 90 $^{o}$C at a constant ramp rate of 0.4K/s, the R-T curve for the H-$VO_2$ (I) insulator film shows a gradual transformation to the metallic H-$VO_2$ (M) state. (d) The metallic H-$VO_2$ (M) reverts to the monoclinic insulator state after baking in air at 180$^{o}$C. After the 30min treatment, recovery to the initial M-$VO_2$ state is nearly complete.

FIG. 2 (a) XPS spectra for $VO_2$ films: the as-prepared M-$VO_2$ film, hydrogenated metallic (M) and insulator (I) films, and dehydrogenated film, showing clear differences between the V2p and O1s peaks. (b) The UV-Vis spectra for hydrogenated samples during phase transitions. (c) The crystal lattice variations during the hydrogenation process. (d) The synchrotron based O K-edge XANES spectra mapping for the electron occupancy of the $t_{2g}$ and $e_g$ orbitals during the hydrogenation process.

FIG. 3 Optimized atomic structures (top panel), density of states (DOS: middle panel), band structures (bottom panel) of the pure monoclinic $VO_2$ unit cell ($V_4O_8$) and H-doped cells ($H_xV_4O_8$ with x=2, 4). (a), (b), (c) are the crystal structures of $V_4O_8$, $H_2V_4O_8$, and $H_4V_4O_8$, respectively. The red, grey, and blue balls represent O, V and H atoms, respectively. (d), (e), (f) are the DOS of $V_4O_8$, $H_2V_4O_8$ and $H_4V_4O_8$, respectively, where the Fermi level $E_F$ is marked with dashed lines. (g), (h), (i) are the band structures of $V_4O_8$, $H_2V_4O_8$ and $H_4V_4O_8$, respectively.

FIG. 4 (a) The DOS of $H_3V_4O_8$, in which the shaded areas represent occupancy of the conduction band edge state, corresponding to DOS below the Fermi level $E_F$. (b) The electron occupancy of the total and V-3d DOS (computed by integrating the shaded areas) as a function of H-doping concentration.

*Figure-1*

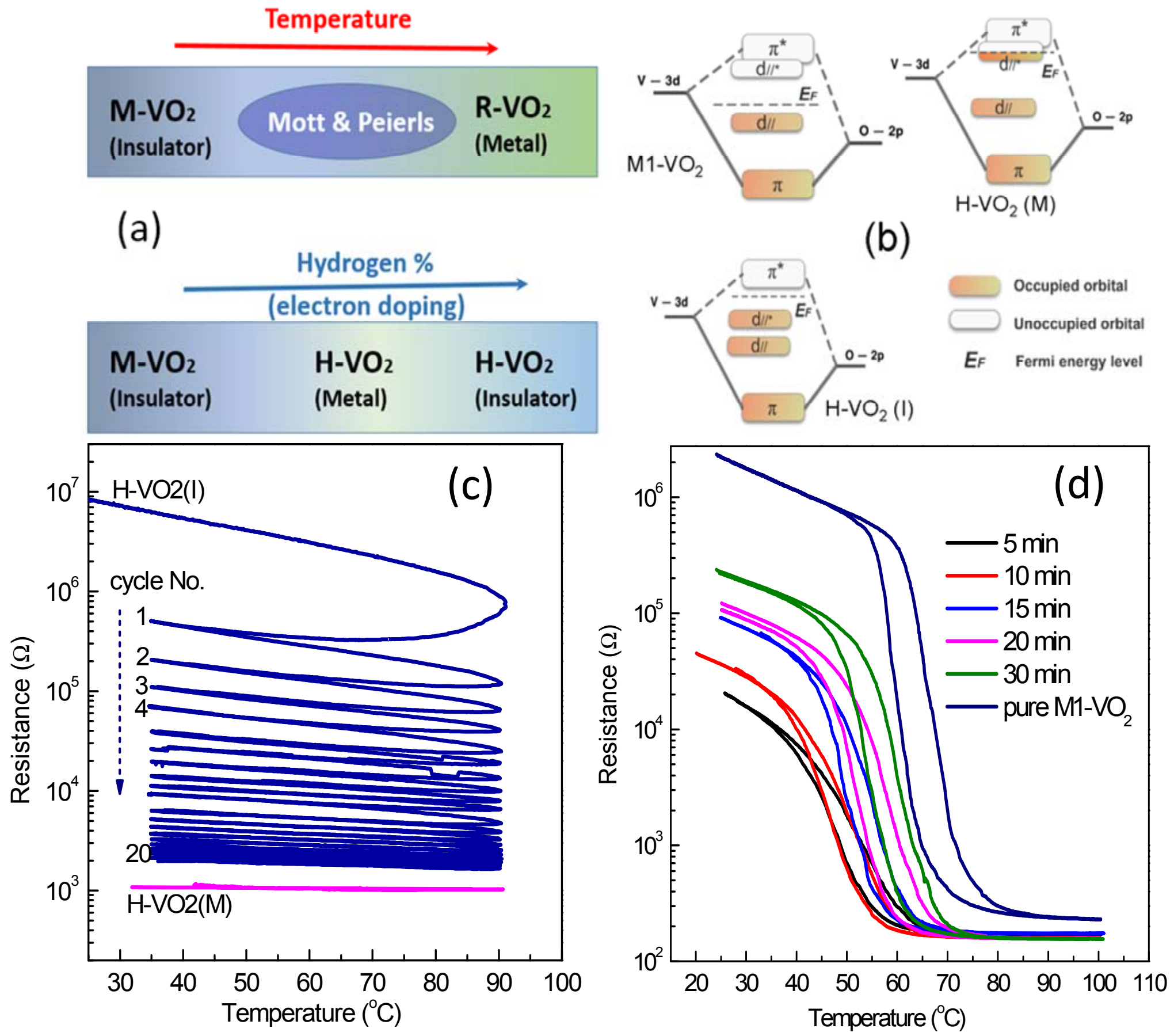

*Figure-2*

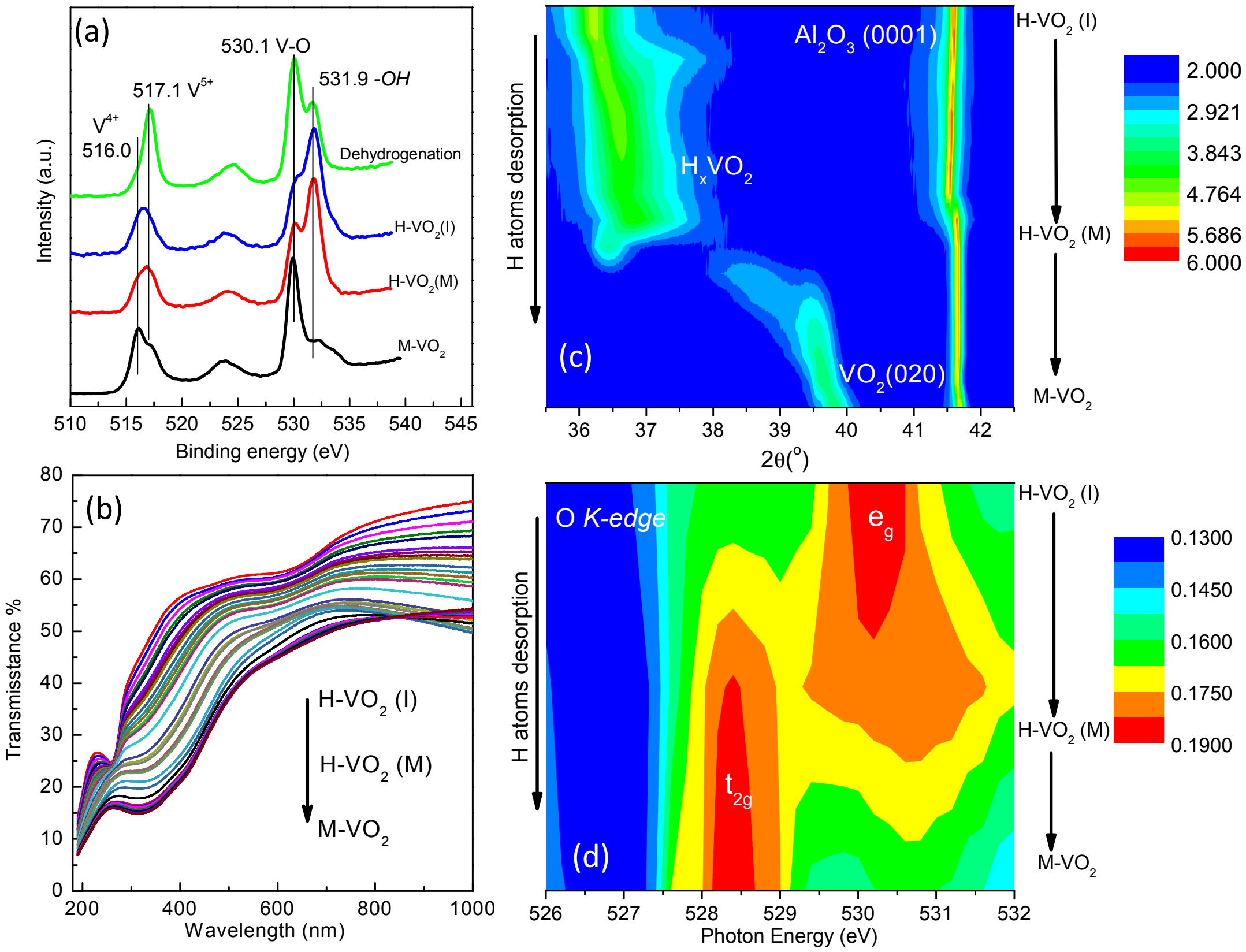

*Figure-3*

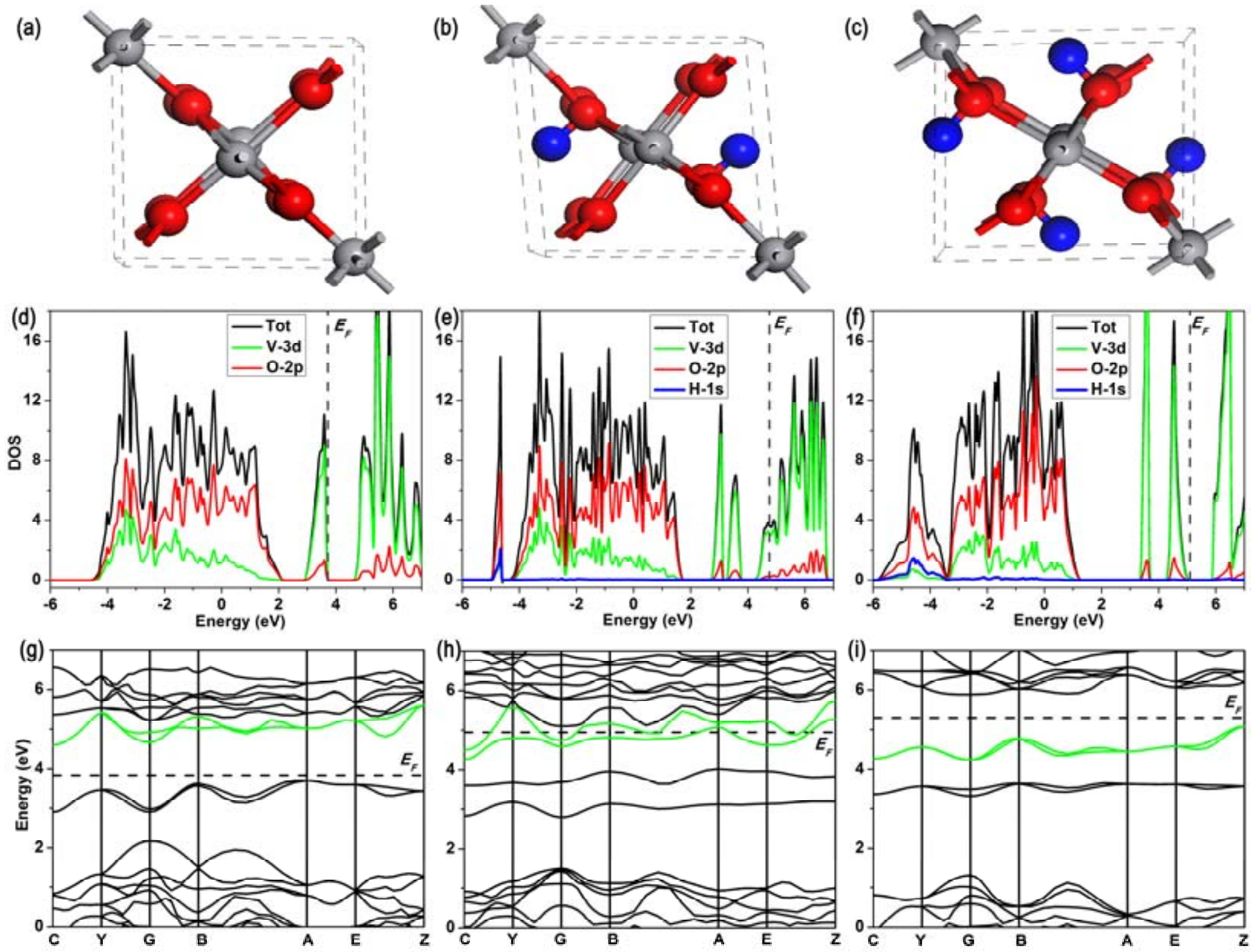

*Figure-4*

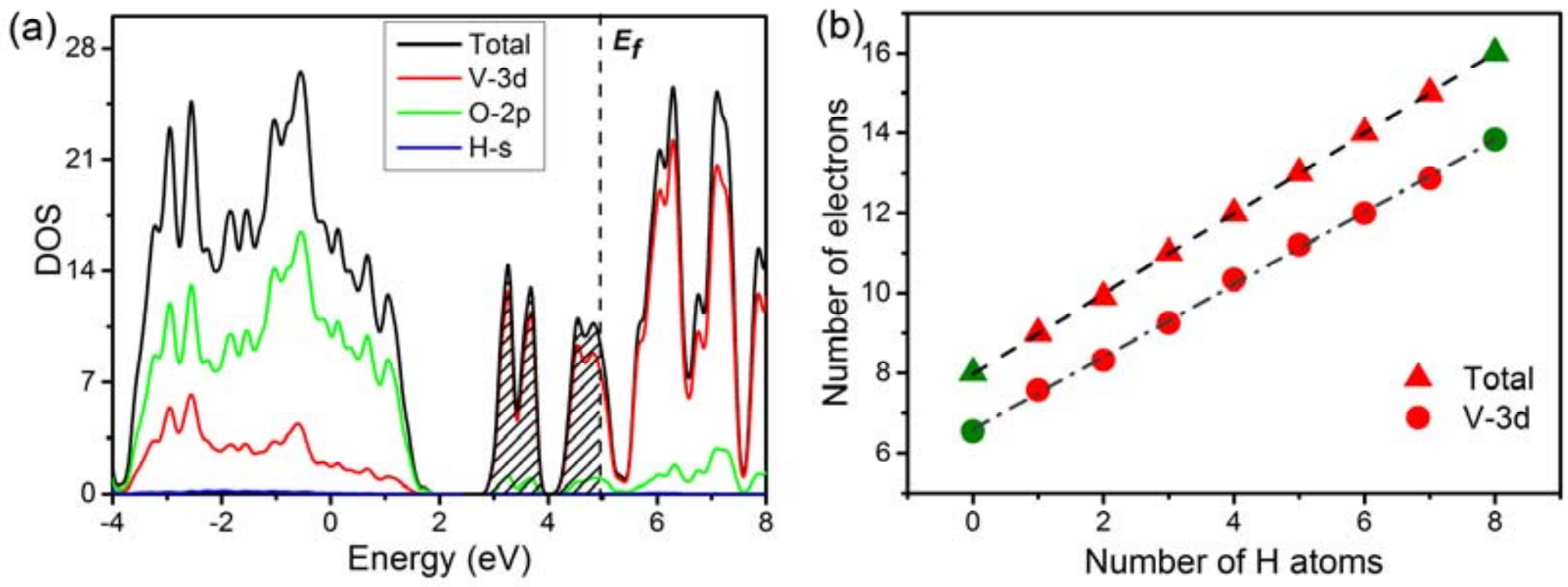